\documentclass{andp2012}
\usepackage[english]{babel}
\usepackage{url}
\title{State of the art in the determination of the fine structure constant: test of Quantum Electrodynamics  and determination of $h/m_{\mathrm{u}}$} 
\author[R. Bouchendira]{R. Bouchendira\inst{1}}
\author[P. Clad\'e]{P. Clad\'e\inst{1}}
\author[S. Guellati-Kh\'elifa]{S. Guellati-Kh\'elifa\inst{1,2,}\footnote{S. Guellati-kh\'elifa\quad E-mail:~\textsf{guellati@spectro.jussieu.fr}}}
\author[F. Nez]{F. Nez\inst{1}}
\author[F. Biraben]{F. Biraben\inst{1}}

\address[1]{Laboratoire Kastler Brossel, Ecole Normale Sup\'erieure, Universit\'e Pierre et Marie Curie, CNRS, 4 place Jussieu, 75252 Paris Cedex 05, France}
\address[2]{Conservatoire National des Arts et M\'etiers,
292 rue Saint Martin, 75141 Paris Cedex 03, France}
\shortauthors{R. Bouchendira et al.}
\begin{abstract}
The fine structure constant $\alpha$ has a particular status in physics. Its precise determination is required to test the quantum electrodynamics (QED) theory. The constant $\alpha$ is  also a keystone for the determination of other fundamental physical constants, especially the ones involved in the framework of the future International System of units.
This paper presents Paris experiment, where the fine structure constant is determined by measuring the recoil velocity of a rubidium atom when it absorbs a photon. The impact of the recent improvement of QED calculations of the electron moment anomaly and the recent measurement of the cesium atom recoil at Berkeley will be discussed. The opportunity to provide a precise value of the ratio $h/m_{\mathrm{u}}$ between the Planck constant and the  atomic mass constant will be investigated.
 \end{abstract}
\shortabstract
\begin{document}
\maketitle
\section{Introduction}
Since its discovery at the beginning of the 20$^{th}$ century up to nowadays, the fine structure constant $\alpha$ remains one of the most fascinating fundamental constants,  as it is dimensionless. Currently it plays a central role in the Physics of the 21$^{st}$ century by testing the most accurate theories such as  quantum electrodynamics (QED) \cite{Bouchendira2011, Hanneke2008, Aoyama2012}, testing the stability of fundamental constants ($\dot{\alpha}/\alpha$) (for example see review by J.P. Uzan \cite{Uzan}) but also in a practical way in the proposed redefinition of the international system of units (SI) \cite{Mills2011}.

The name of the fine structure constant derives from the Sommerfeld model \cite{Sommerfeld1916}. It was intended to explain the fine structure of the hydrogen spectral lines, unaccounted for in the Bohr model. The Sommerfeld model combines the theory of relativity with the Bohr model. The constant $\alpha$ appears in the velocity of the electron ($v_\mathrm{e}$) on its first orbit around the proton ($v_\mathrm{e}=\alpha\times c$, where $c$ is the velocity of light). The expression for $\alpha$ is:
\begin{equation}
\alpha=\frac{e^2}{4 \pi \epsilon_{0} \hbar c}\label{eqalpha}
\end{equation}
where $e$ is the charge of the electron, $\epsilon_{0}$ the vacuum permittivity and $\hbar=h/2\pi$ in which $h$ is the Planck constant.

The Sommerfeld model failed because it didn't take into account the spin of the electron. Nevertheless the constant introduced in this model is still relevant in the Dirac model which combines  relativity and  quantum mechanics \cite{Dirac1916}. This model predicts the existence of the positron and the spin of the electron! In 1947 a new effect from which the value of $\alpha$ can be deduced was discovered: the vacuum quantum fluctuations which contribute to the splitting of $2S_{1/2}$ and $2P_{1/2}$ energy levels in hydrogen (now usually called the Lamb shift) \cite{Lamb1947, Lamb1950} and also  contribute to the anomaly of the gyromagnetic factor of leptons \cite{Schwinger1948, Kusch1948}.

Indeed the modern understanding of $\alpha$ is that it sets the scale of the electromagnetic interaction. Consequently many experiments in which a charged particle interacts with an electromagnetic field can be used to determine $\alpha$. In  1998, the experiments considered by the CODATA task group on fundamental constants to give the best estimate of the fine structure constant value ranged from solid state physics and atomic physics to quantum electrodynamics \cite{CODATA98}.

\begin{figure}
\includegraphics[width=\columnwidth]{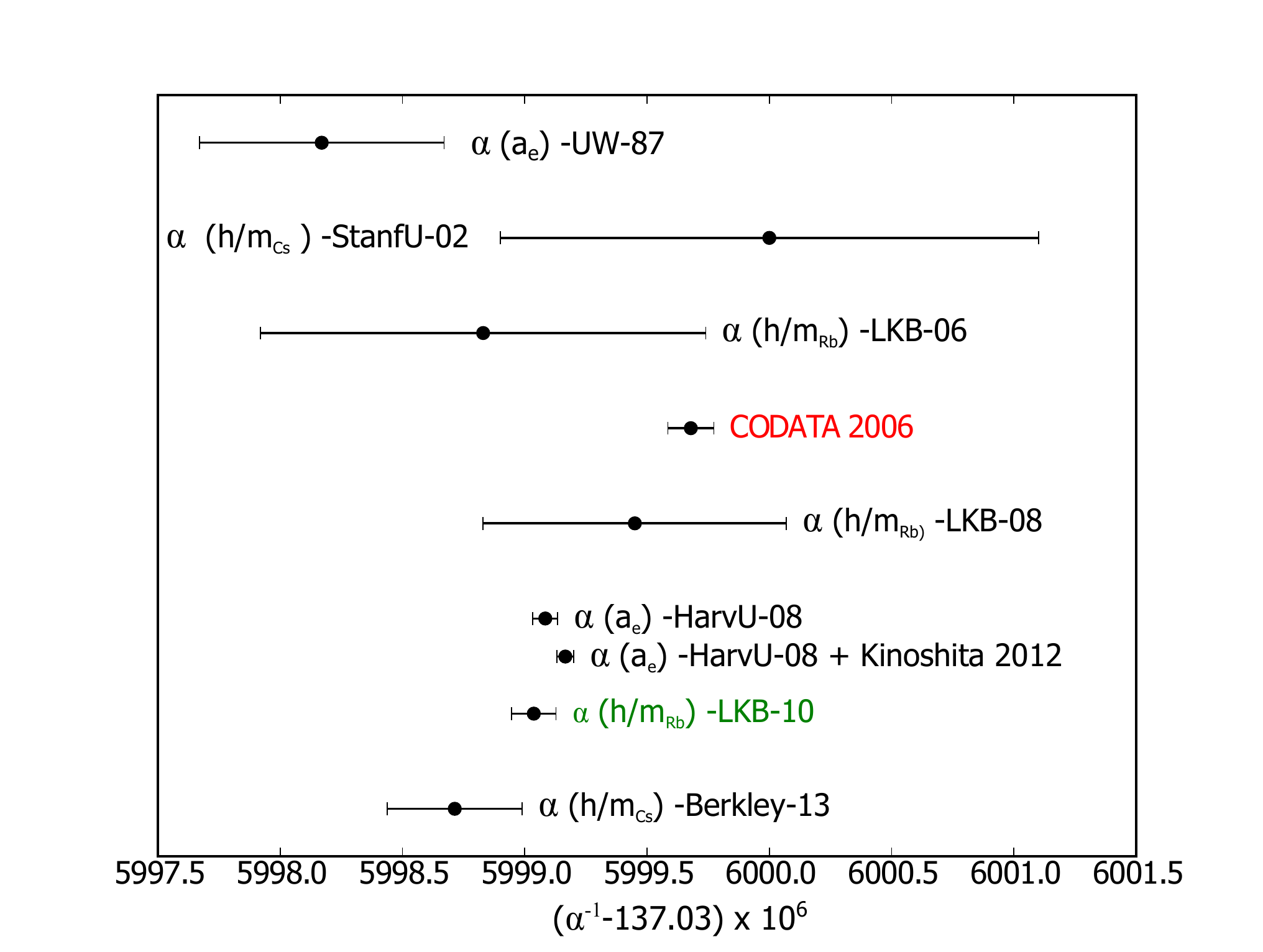}%
  \caption{\label{Figure.1}\col
    The most precise determinations of the fine structure constant. The value labelled $\alpha(h/m_{\mathrm{Cs}})$ is calculated using the value of the Compton frequency of cesium atoms measured by Muller's group \cite{Muller2013}. The values labelled ($\alpha(a_{\mathrm{e}})-\mathrm{UW-87}$) and ($\alpha(a_{\mathrm{e}})-\mathrm{HarvU}$) are deduced from the experimental values of the electron moment anomaly $a_{\mathrm{e}}$ performed respectively by Dehmelt at the University of Washington \cite{VanDick} and Gabrielse at Harvard university \cite{Hanneke2008}.}
\end{figure}

As shown in figure~\ref{Figure.1}, the current most precise determination of the fine structure constant comes mainly from two methods. 

The first method combines the measurement of the electron magnetic
moment anomaly $a_{\mathrm{e}}$ and QED perturbation theory. The value of $\alpha$ is determined by comparing the experimental value of $a_{\mathrm{e}}$ with :
\begin{align}
    a_{\mathrm{e}} &=&A_1\times\frac{\alpha}{\pi}+A_2\times\left(\frac{\alpha}{\pi}\right)^2+A_3\times\left(\frac{\alpha}{\pi}\right)^3+A_4\times\left(\frac{\alpha}{\pi}\right)^4+.....\\
    &&+a_{\mathrm{e}}(\frac{m_\mathrm{e}}{m_{\mu}},\frac{m_\mathrm{e}}{m_{\tau}},\mathrm{weak}, \mathrm{hadron})\label{eqQED}
\end{align}
Thus in the QED model $a_{\mathrm{e}}$ is expressed as a power series of $\alpha$ and an additive term  which takes into account the contributions due to the muon, the tau, the weak and hadronic interactions. The coefficients $A_i$ are finite and dimensionless constants calculated by using Feynman diagrams \cite{Aoyama2012}. The last term includes the Lepton mass dependence, the weak and hadronic interactions.

The second one, introduced by the group of S. Chu at Stanford university \cite{Wicht:02}, is based on  the measurement of the ratio $h/m_{\mathrm{X}}$, between the Planck constant $h$ and the atomic  mass $m_\mathrm{X}$. This ratio is related to $\alpha$ by
\begin{equation}
       \alpha^2=\frac{2R_\infty}{c}\frac{A_r(\mathrm{X})}{A_r(\mathrm{e})}\frac{h}{m_{\mathrm{X}}}\label{calculalpha}
\end{equation}
The Rydberg constant $R_\infty$ is  known with an accuracy of $5 \times 10^{-12}$ \cite{CODATA2006,Udem1997,Schwob1999}. 
 The uncertainty on the relative mass of  the electron $A_r(\mathrm{e})$ and the relative atomic mass $A_r(\mathrm{X})$ are respectively $4.4 \times 10^{-10}$\cite{CODATA12} and less than $10^{-10}$ for Rb and Cs \cite{Bradley1999, Mount2010}. Using an atom interferometer and Bloch oscillations, we have performed in 2010 a determination of the ratio $h/m_{\mathrm{Rb}}$. The value $\alpha(h/{m_{\mathrm{Rb}}})$ that has been deduced is the most precise value obtained using this method\cite{Bouchendira2011}.

The comparison of these two determinations is one of the most precise tests of QED. It is so accurate that one can think, in a near future, of using these lab-size experiments to check theoretical predictions tested up to now only on particle accelerators (for example the existence of internal structure of the electron \cite{Gabrielse2006}).

For many years, the main contribution to the determination of $\alpha_\mathrm{CODATA}$ has been the one derived from the anomaly of the gyromagnetic factor of the electron ($\alpha(\mathrm{a_{e}})$) which is strongly dependent on complex QED calculations.
Nowadays the uncertainties of $\alpha(\mathrm{a_{e}})$  and $\alpha(\mathrm{Rb})$ are in the same order of magnitude. This makes the CODATA adjustment more reliable.

This reliability is essential for the redefinition of the SI which will rely on the values of fundamental constants \cite{CGPM2011, Borde2005,Mills2006}. In the proposed redefinition, the Planck constant will have a fixed value in SI units\cite{Mills2011}. In order to link the microscopic definition to the macroscopic Kilogram, two kinds of experiments are competitive. The first one, the watt balance measures the ratio $h/M$ between the Planck constant and a macroscopic standard mass $M$ \cite{Kibble1990, Steiner2007,Williams1998}. In the current SI, it gives a determination of $h$. In the future SI, it will give the measurement of a macroscopic mass. The second experiment is the Avogadro project, which directly determines the ratio $M/m$ between a macroscopic mass (the mass of a silicon sphere) and a microscopic mass (the mass of the atom of silicon) \cite{Andreas2011}. In the current SI, it gives a determination of the (unified) atomic mass constant $m_{\mathrm{u}}$ defined according to $m_\mathrm{u}$ = $m(\mathrm{^{12}C})$/12, or the Avogadro constant. 
The ratio $h/m_{\mathrm{u}}$ provides therefore a direct comparison between the two experiments. Its precise determination has a major interest in metrology.
Whereas the photon-recoil measurement, combined with the appropriate relative atomic mass measurement, gives a determination of the ratio $h/m_\mathrm{u}$, other values of $\alpha$ can be converted into $h/m_\mathrm{u}$ using the formula:
\begin{equation}
\frac h{m_\mathrm{u}} = \frac{\alpha^2 c A_r(\mathrm{e})}{2R_\infty}\label{mu}
\end{equation}

We emphasize that the ratio $h/m_{\mathrm{u}}$ becomes identified with Avogadro Planck constant as:
\begin{equation}
N_{\mathrm{A}}h=\frac {h}{m_\mathrm{u}}\frac {M(^{12}\mathrm{C})}{12}\label{NAh}
\end{equation}
where M($^{12}$C) =12$\times10^{-3}$ kg/mol is the carbon molar mass and $N_{\mathrm{A}}$ is the Avogadro constant. The product $h N_{\mathrm{A}}$ is in the current SI, equivalent to the ratio $h/m_\mathrm{u}$. 
It seems to us more relevant to consider $h/m_{\mathrm{u}}$ in the framework of the redefinition of the kilogram. In the future SI of units, the Avogadro constant $N_{\mathrm{A}}$, which is used by the chemists to quantify and identify an amount of substance with atoms and molecules, will be fixed. This will break the link between atomic masses and molar masses. Consequently $M(^{12}$C) will no longer be equal to 12 g/mol, but will be determined from equation \ref{NAh} using the ratio $h/m_{\mathrm{u}}$.

In the proposed new International Systems of Units, many others physical constants, that are set by the CODATA will have a fixed value. The constant $\alpha$ will be a keystone of the proposed SI, as many of the remaining constants will depend strongly on its knowledge (such as the vacuum permeability $\mu_0$, the von Klitzing constant $R_\mathrm{K}$, ...)\cite{Mills2011}.

\medskip

The next and largest section of this paper will be devoted to the  experiment in Paris. This experiment started in 1998 and was entirely renewed in 2008. In the last part, we will discuss the role of the various determinations of $\alpha$. We will focus on the test of QED calculations and on the impact on the redefinition of the Kilogram.

\section{Determination of the ratio $h/m_{\mathrm{Rb}}$}
\subsection{Principle}
The ratio $h/m_{\mathrm{Rb}}$ is deduced from the measurement of the recoil velocity $v_r$ of an atom when it absorbs a photon ($v_r=\hbar k/m$ with $\hbar$ the reduced Planck constant, $k$ the wave vector and $m$ the mass of atoms. This measurement is performed by combining a Ramsey-Bord\'e atom interferometer \cite{Borde1989} with the Bloch oscillations technique.  Bloch oscillations have been first observed in atomic
physics by the groups of Salomon and Raizen \cite{BenDahan,Peik,Wilkinson1996}. The
atoms are shed with two counter-propagating laser beams whose frequency difference
is swept linearly. One can then consider that the atoms are
placed in a standing wave which is accelerated when the
frequency difference between the two laser beams is swept.
The atoms experiment an inertial force in a periodic optical potential.
This system is analogous to the BO of an electron in a solid
submitted to an electric field. Another point of view is to consider that 
the atoms undergo a succession
of Raman transitions which correspond to the absorption of
one photon from a beam and a stimulated emission of another
photon to the other beam. The internal state is unchanged
while the atomic velocity increases by 2$\times v_r$ per oscillation.
The Doppler shift due to this velocity variation is periodically
compensated by the frequency sweep and the atoms
are accelerated. For 87-rubidium atoms the Doppler shift induced by a variation of velocity  of 2$\times v_r$ is 30 kHz, the number of Bloch oscillations performed by the atoms is set precisely by the frequency sweep. 
In our previous work we demonstrated that BO is a very efficient process in terms of photon momentum transfer\cite{Battesti04}.

\begin{figure}
  \includegraphics[width=\columnwidth]{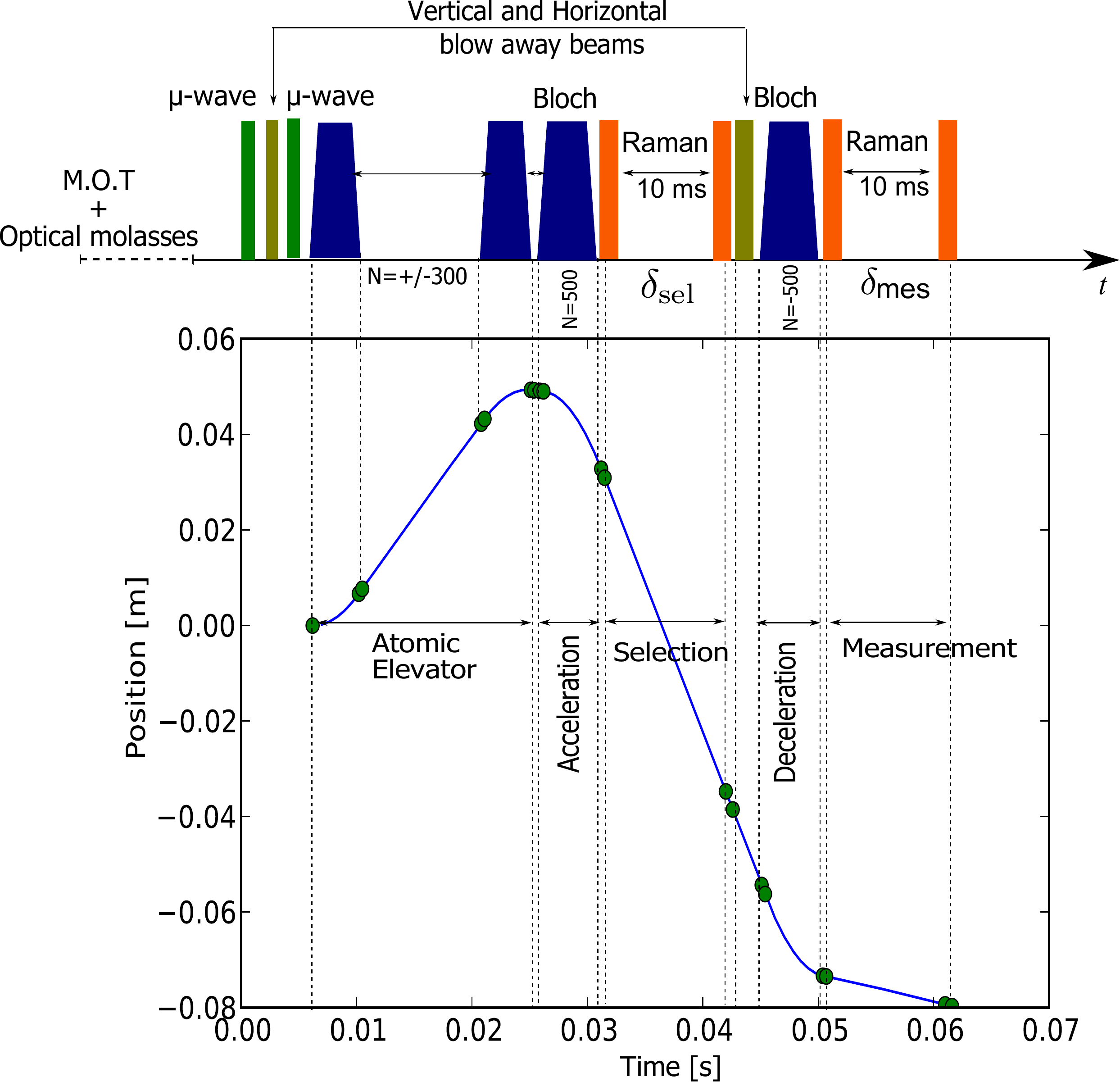}%
  \caption{\label{Figure.2}\col
     The pulses timing sequence and atomic trajectory during the measurement procedure.}
\end{figure}

The timing sequence of the experiment is depicted in figure~\ref{Figure.2}. The 87-rubidium atoms are first cooled in a magneto-optical trap and optical molasses in the $F=2$ hyperfine level. A $\mu$-wave excitation is used to select atoms in $F=2, m_F=0$ : we apply a vertical magnetic field of 7 $\mu$T, a first $\mu$-wave excitation pulse transfers atoms from $F=2, m_F=0$ to $F=1, m_F=0$  Zeeman sub-level. The blow-away laser beam cleans the $F=2$ hyperfine level. The atoms in  $F=1, m_F=0$ are returned to $F=2, m_F=0$ using a second $\mu$-wave pulse. In order to increase the interaction area, we apply an atomic elevator to displace the atoms toward the lower or the upper side of the vacuum chamber. The atoms are accelerated and then decelerated by the means of two Bloch pulses delayed by 10.3 ms. Each one transfers to the atoms 600 $\times v_r$ during 4.6 ms in a given direction. The atomic elevator carries 30$\%$ of the atoms, which represents the whole proportion of the atoms which fit in the first Brillouin zone. 

We  then start  the measurement procedure by accelerating the atoms with 500 BO in 5.6 ms (for details see \cite{Cadoret2008, Clade2006}). The velocity of the atoms is measured by using a Ramsey-Bord\'e atom  interferometer performed by two pairs of $\pi/2$ pulses. The delay $T_\mathrm{R}$ between two $\pi/2$ pulses is 10 ms and the duration of each pulse is $\tau=600~\mu s$. The laser pulses induce a Doppler sensitive Raman transition between the hyperfine levels $F=2$ and $F=1$, thus the velocity is measured in terms of frequency. The first pair of $\pi/2$ pulses transfers the resonant velocity class from $F=2$ to $F=1$. We then shine a resonant laser beam ($F = 2\longrightarrow F' = 3$ transition) to push away atoms remaining  in $F=2$ before coherently accelerating  atoms in $F=1$ with 500 BO. The final velocity of the accelerated atoms is measured with a second pair of $\pi/2$ pulses by transferring atoms from $F=1$ to $F=2$. The population in each hyperfine level is measured with a time of flight technique.
The fringe pattern which represents the final velocity distribution is obtained by scanning the frequency of the Raman lasers during the final $\pi/2$  pulses.

\subsection{Experimental setup}

\begin{figure}
\includegraphics[width=\columnwidth]{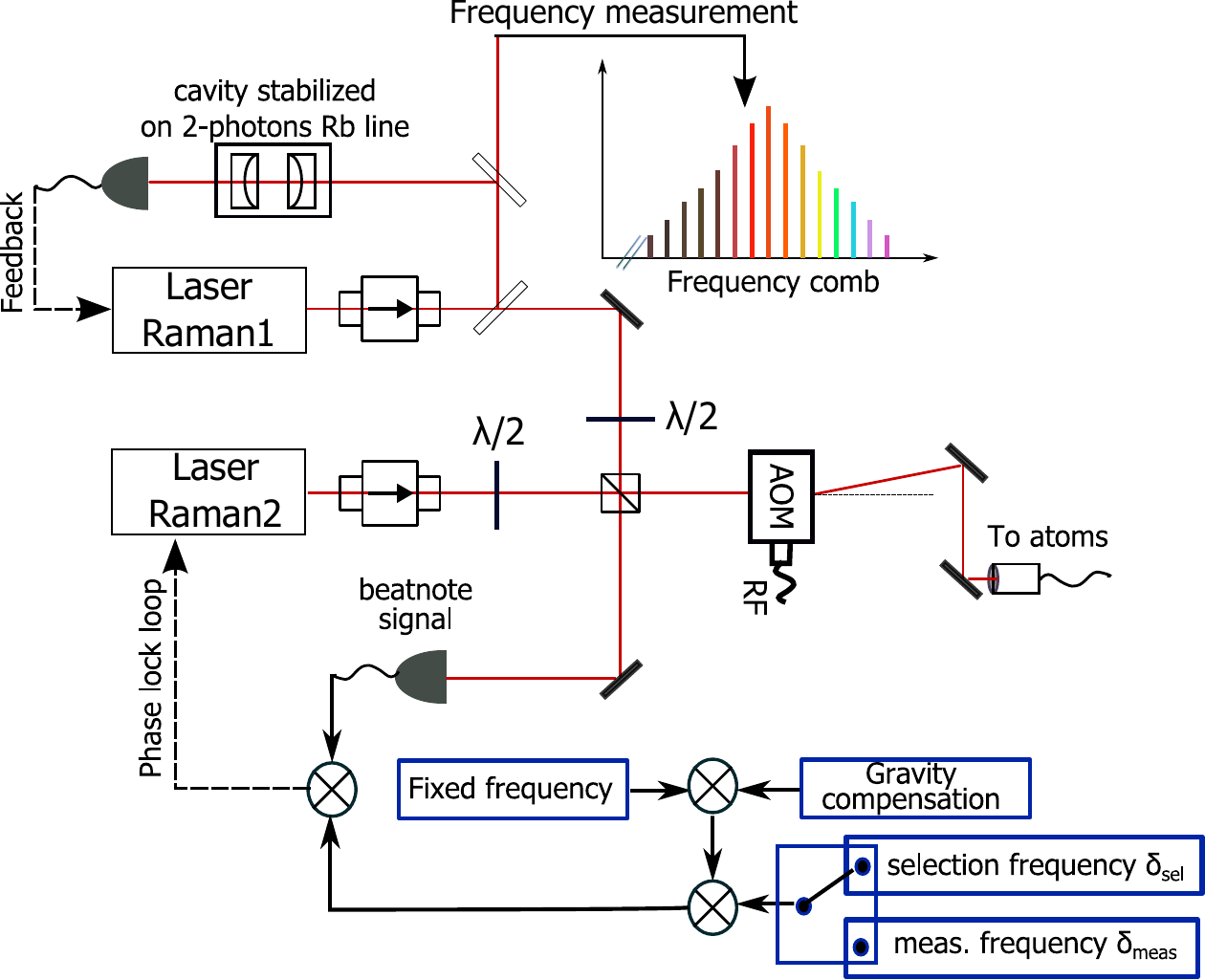}%
  \caption{\label{Figure.3}\col
   The optical setup of the Raman beam used to perform the atomic interferometer. The two laser diodes are stabilized using an interference-filter-stabilized extended cavity. They are phase locked. The frequency of one Raman laser is stabilized on an ultra-stable cavity and measured with a femtosecond comb.}
\end{figure}

A two-dimensional magneto-optical trap (2D-MOT) produces a
slow atomic beam (about 10$^9$ atoms/s at a velocity of
20 m/s) which loads during 250 ms a three-dimensional
magneto-optical trap. Then a $\sigma^+$-$\sigma^-$ molasses generates
a cloud of about 2 $\times$ 10$^8$ atoms in the $F =2$ hyperfine
level, with a 1.7 mm radius and at a temperature of 4 $\mu$K. The 2D-MOT cell is a glass cell separated from a UHV-chamber by a differential pumping tube which is also the aperture for the output slow beam. The cooling and pumping lasers are interference-filter-stabilized external-cavity diode lasers (IF-ECL) \cite{Baillard2006}, both lasers are amplified in the same tapered amplifier. The frequency of the cooling beam is actively controlled by using the beatnote signal with the pumping beam, itself locked on a suitable rubidium crossover line.  
The Raman lasers are also IF-ECL diode lasers. The two diode lasers are phase-locked using a synthesized frequency referenced to a cesium atomic clock. As shown in figure~\ref{Figure.3}, the synthesized frequency results from a mixing of a fixed frequency (6.84 GHz), a frequency ramp  to compensate the fall of atoms in the gravity field (25 kHz/s) and the probe frequency. The probe frequency is switched between $\delta_\mathrm{sel}$ and  $\delta_\mathrm{meas}$ using two independent synthesizers, where $\delta_\mathrm{sel}$ and $\delta_\mathrm{meas}$ are the frequency differences between the two Raman beams respectively during the first and the second pairs of $\pi/2$ pulses of the atom interferometer. 
The Raman beams are  blue-detuned by 125 GHz from the 87-Rubidium D2 line. The Bloch beams originate from a 2.5 W Ti:sapphire laser. The output laser beam is split into two paths, each of which passes
through an AOM to adjust the frequency offset and amplitude before being injected into a polarization maintaining fibre. The depth of the generated optical lattice is 45$E_r$ ($E_r$ is the recoil energy) for an effective power of 150~mW seen by the atoms. The optical scheme of the Bloch and the Raman beams is described in detail in \cite{Bouchendira2011,Cadoret2008}. The frequencies of one Raman laser and the Bloch laser are stabilized onto a same ultra-stable Zerodur Fabry-Perot cavity, itself stabilized on the $5S_{1/2}(F=3)\longmapsto 5D_{3/2} (F=5)$ two-photon transition of 85-rubidium\cite{Touahri} (short term). On the long term, these frequencies are precisely measured by using a femtosecond  comb  referenced to the cesium clock. As the measurement of the ratio $h/m_{\mathrm{Rb}}$ is performed in terms of frequency, it is thus directly connected to the cesium standard.

The vacuum chamber is supported by an active vibration isolation platform. The residual acceleration is reduced by a factor of 100 above 10 Hz. However, vibrations remain one of the main limitations in the sensitivity of the atom interferometer.

\subsection{Results}

\begin{figure}
\includegraphics[width=0.8\columnwidth]{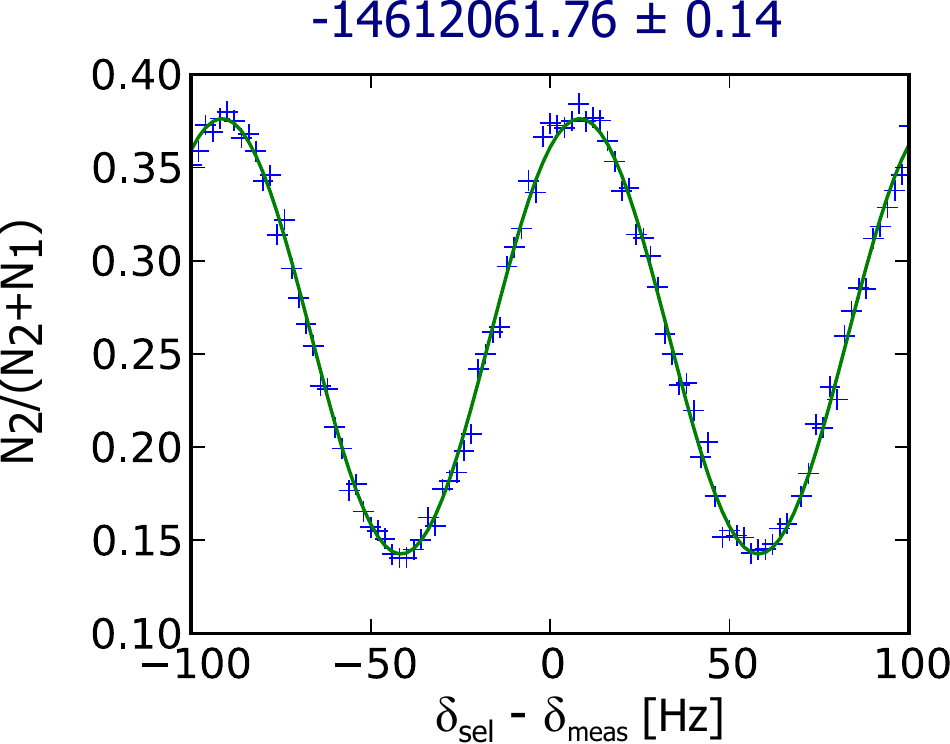}%
  \caption{\label{Figure.4}\col
    The quantity $N_2/(N_2+N_1)$ versus  the frequency difference
between the two pairs of $\pi/2$ pulses, where $N_2$ and $N_1$ represent respectively the populations in hyperfine levels $F=1$ and $F=2$. The spectrum is recorded with 100 points during 1 min. The measured position of
the central fringe is indicated above the spectrum.}
\end{figure}

In figure~\ref{Figure.4} we show a typical fringe pattern obtained  with 100 points during 1 min.  The central fringe is determined with an uncertainty  of 0.14 Hz  corresponding to the relative uncertainty of 10$^{-8}$ on the Doppler shift $(\delta_\mathrm{sel}-\delta_\mathrm{meas})$ induced by 500 BO.

A value of $h/m_{\mathrm{Rb}}$ is obtained by recording four spectra obtained under different conditions to cancel systematic errors and then using:

\begin{equation}
\frac{\hbar}{m_{\mathrm{Rb}}}=\frac{1}{4}\sum_{Spectra}\frac{2\pi\left\vert \delta_\mathrm{sel}-\delta_\mathrm{meas}\right\vert }{2 N k_\mathrm{B}(k_1+k_2)}
\end{equation}
where $k_1$ and $k_2$ are the wave-vectors of the Raman laser beam, $k_\mathrm{B}$ is the wave-vector of the Bloch laser beams and $N$ the number of Bloch oscillations.

\begin{table*}
\centering
  \begin{andptabular}[\textwidth]{lcccc}{Typical experimental parameters for the determination of one value of $h/m_{\mathrm{Rb}}$\label{table1}}%
$N_{\mathrm{elev}}$(1)& -300&-300&+300&+300\\
$N_{\mathrm{elev}}$(2)& +300&+300&-300&-300\\
$N_{\mathrm{up}}$& +500&+500&-500&-500\\
$N_{\mathrm{down}}$& -500&-500&+500&+500\\
Raman beams direction& +1&-1&+1&-1\\
$(\delta_{sel}-\delta_{meas})$(Hz)&15567824.42&-15567822.07&-14612062.24&14612067.77\\
Uncertainty on the central fringe& 0.15&0.16&0.13&0.16\\
  \end{andptabular}
\end{table*}

Two spectra allow to get rid of the change in velocity due to the free fall of atoms in the gravity field. They are obtained by accelerating the atoms alternatively upward and downward. The difference between the results eliminate
 $gT$, where $T$ is the spacing time between the two pairs of Raman $\pi/2$ pulses.  Otherwise for each initial  acceleration, we record two other spectra by exchanging the direction of the Raman beams ($\overrightarrow{k_1}$ and $\overrightarrow{k_2}$) in order to cancel the parasitic level shifts due to the Zeeman
effect and the light shifts.
Typically a set of 4 spectra is obtained with the parameters shown on table~\ref{table1}. In this table, $N_{\mathrm{elev}}$(1) and $N_{\mathrm{elev}}$(2) represent the number of BO used to perform the atomic elevator (we first accelerate the atoms with $N_{\mathrm{elev}}(1)$ BO then we stop them using $N_{\mathrm{elev}}(2)$ BO). $N_{\mathrm{up}}$ and $N_{\mathrm{down}}$ are respectively the number of BO for the upward and downward acceleration.  The two last lines of this table give the result of the fit of the central fringe and the corresponding uncertainty. 
Each column gives the parameters for one spectrum. We deduced the  value of $h/m_{\mathrm{Rb}}$ with a relative uncertainty of 5$\times$ 10$^{-9}$ (2.5$\times$ 10$^{-9}$ on $\alpha$).

\begin{figure}
\includegraphics[width=\columnwidth]{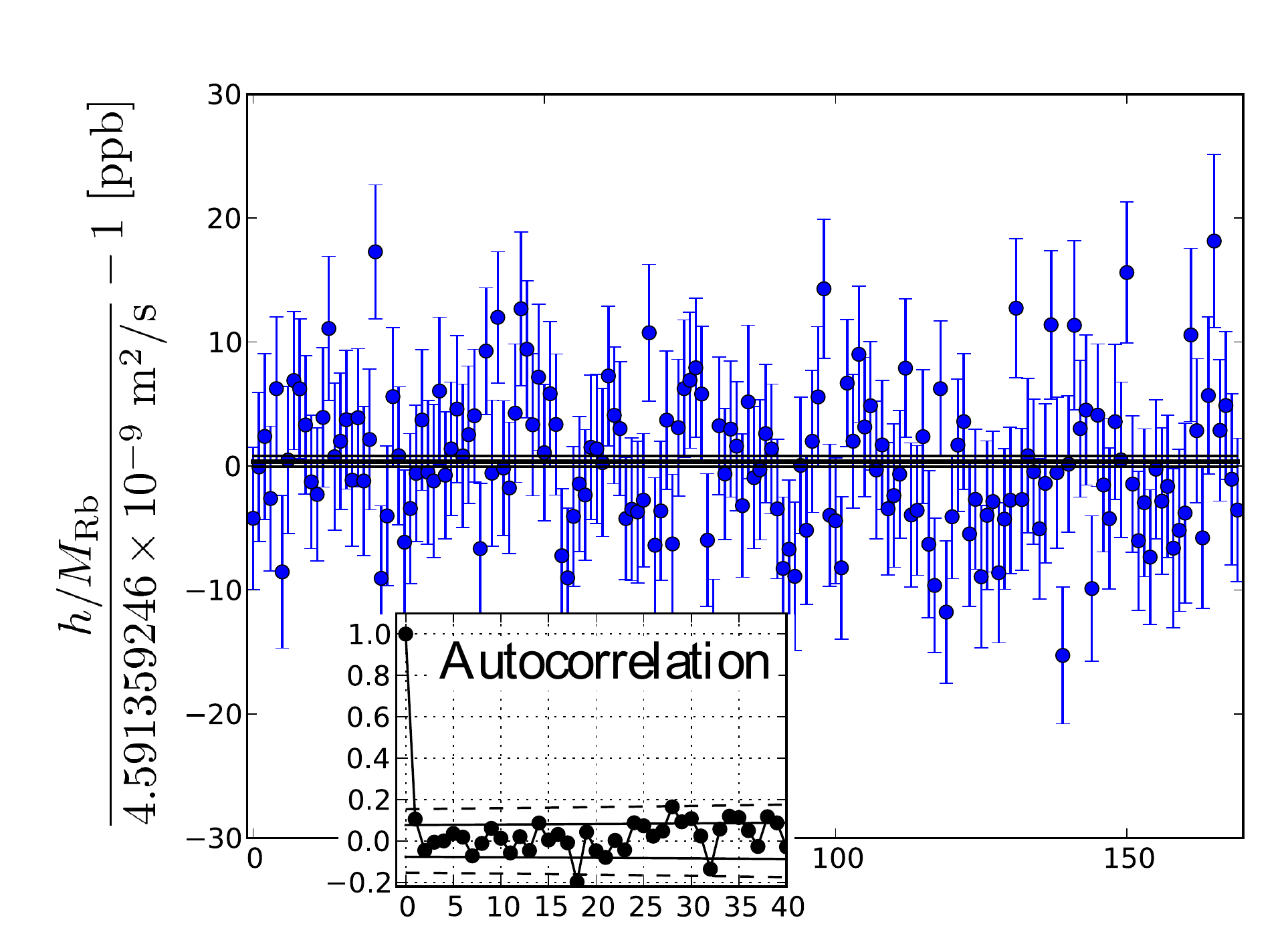}%
  \caption{\label{Figure.5}\col
A selection of 170 measurements of the ratio $h/m_{\mathrm{Rb}}$ obtained during 15 hours integration time. The inset shows the autocorrelation function of these measurements. The solid and the dashed lines represent  the 1$\sigma$ and 2$\sigma$ standard deviations of the autocorrelation function.}
\end{figure}

Figure~\ref{Figure.5} shows a set of 170 determinations of the ratio $h/m_{\mathrm{Rb}}$ recorded for about 15 hours. The standard deviation on   the mean value is 4.4$\times$10$^{-10}$, with a $ \chi^2/(n-1)$=1.05. We have evaluated the autocorrelation function using the approach  described in  reference \cite{Witt2007}. The result is reported in the inset, there is no  correlation between the successive measurements.

\subsection{Systematic effects}
The systematic effects are summarized in the table~\ref{table2}. 
\begin{table}
  \begin{andptabular}[0.8\textwidth]{lrr}{Systematic effects and relative uncertainty in part per 10$^{10}$ on the determination of $\alpha^{-1}$\label{table2}}%
Laser frequencies& &1.3\\
Beams alignment&-3.3& 3.3\\
Wave front curvature and Gouy phase&-25.1 & 3.0\\
2nd order Zeeman effect&4.0 & 3.0 \\
Gravity gradient&-2.0& $0.2$ \\
Light shift (one photon transition)& & 0.1\\
Light shift (two photon transition)& & 0.01 \\
Light shift (Bloch oscillation)& & 0.5 \\
Index of refraction atomic cloud & & \\
and atom interactions& &2.0 \\ \hline
Global systematic effects&-26.4 & 5.9\\ \hline 
Statistical uncertainty& & 2.0\\
Rydberg constant and mass ratio & &2.2 \\ \hline \hline
Total uncertainty& & 6.6\\
  \end{andptabular}
\end{table}

The main systematic effect comes from the Gaussian profile of the laser beams. The atoms experience an effective wave-vector determined by the gradient of the laser phase along the propagation axis z : 
\begin{equation}
k_{\mathrm{eff}}=\frac{d\phi}{dz}=k-\frac{2}{k}\left[ \frac{1}{w^2}-\frac{r^2}{w^4}+\frac{k^2r^2}{4R^2}\right]  
\end{equation}
where $r$ is the radius of the atomic cloud, $w$ the waist of the laser and $R$, the curvature radius. 

\begin{figure}
\includegraphics[width=\columnwidth]{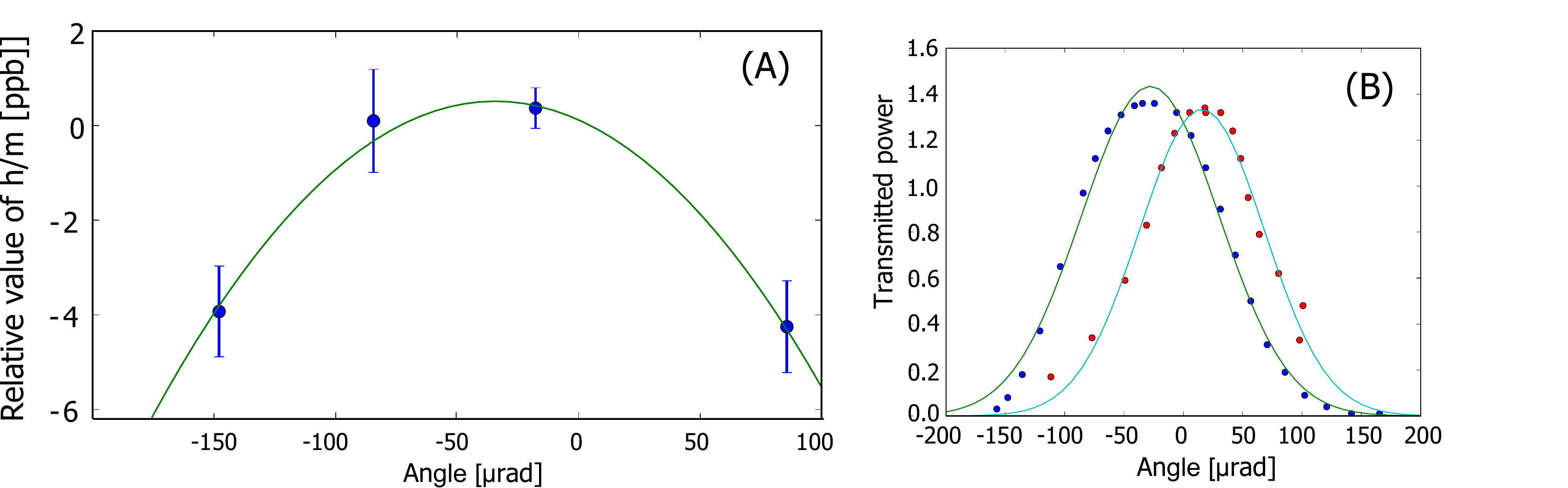}%
  \caption{\label{Figure.6}\col
    The statistical uncertainty on the measurement of the ratio $h/m_{\mathrm{Rb}}$ is sufficient to evaluate the deviation of some experimental parameters. (A): deviation of the misalignment angle between the Bloch beams. (B) Retro-reflection in the optical fibre versus the misalignment angle (green curve : Bloch beams, blue curve : Raman beams).}
\end{figure}

This formula includes both contributions  of the Gouy phase and the wave front curvature.  The geometrical parameters of the laser beams have been carefully measured with a Shack-Hartmann wave-front analyser. The alignment of the laser beams is ensured by controlling the coupling between the two optical fibres. The maximum angle error is estimated to 40 $\mu$rad.  As shown in figure~\ref{Figure.6}-A, this value has been confirmed by considering the deviation of the ratio $h/m_{\mathrm{Rb}}$ versus the angle between the upward and the downward Bloch beams (see figure~\ref{Figure.6}-B). The experimental protocol allows to cancel a large part of the level shifts (Zeeman and light shift). This cancellation is performed in three ways: between the selection and the measurement
Raman pulses, between the upward and downward trajectories,
and when the Raman beams direction is changed. The vacuum chamber is enclosed in a double magnetic shield, we have precisely evaluated the residual magnetic field along the interaction area  using Zeeman sensitive Raman transitions. The correction on $\alpha$ due to the second order Zeeman shift is estimated to 4$\times$10$^{-10}$. The light shift is mainly due to the expansion of the atomic cloud between the selection step and the measurement step and the unbalance of the laser intensity when we exchange the direction of the Raman beams.

The density of the atomic cloud after the $\mu$-wave selection and the elevator sequence is about 2$\times$10$^8$ atoms/cm$^3$. The correction due to the index and the mean-field effects is estimated to 10$^{-10}$ with a conservative uncertainty of 2$\times$10$^{-10}$(see \cite{Clade2006}). Compared to the measurements made in 2008, the effect of the background vapour is now negligible (a few parts per 10$^{-11}$), thanks to the differential pumping in the double-cell design.
Finally the deduced value of the fine structure constant is :
\begin{equation}
\alpha^{-1}(h/m_{\mathrm{Rb}})=137.035999044(90) ~~[0.66 \mathrm{ppb}]
\label{resultat}
\end{equation}
This value is slightly different from that published in \cite{Bouchendira2011}, where we have used the value of the rubidium mass determined by B. J. Mount et al.,\cite{Mount2010}. To obtain the result of equation \ref{resultat}, we have used the mean value between the mass values published in references  \cite{Bradley1999} and \cite{Mount2010}.

\section{Discussion}

\begin{figure}
\includegraphics[width=\columnwidth]{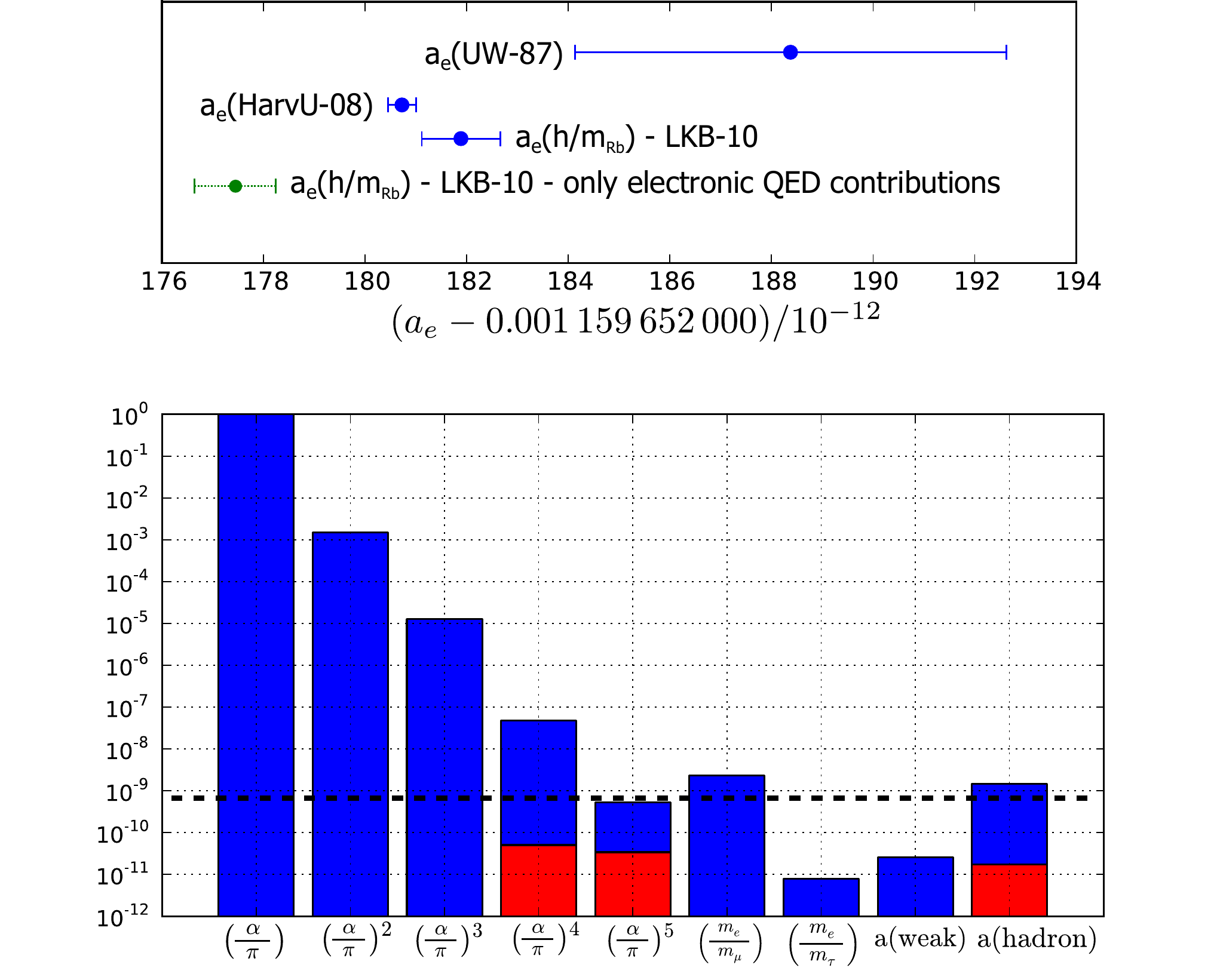}%
  \caption{\label{Figure.7}\col
   Upper figure: comparison of the measurements of the electron
moment anomaly ($a_{\mathrm{e}}(\mathrm{UW-87})$ and $a_{\mathrm{e}}(\mathrm{HarvU-08})$) with the
theoretical value obtained using $\alpha(h/m_{\mathrm{Rb}})$. The green point is obtained without the last term of  equation~\ref{eqQED}. The lower figure in blue, relative contributions to the electron anomaly of the different terms of
equation~\ref{eqQED}, in red their uncertainties. The dashed line
corresponds to the relative uncertainty on the value of $\alpha(h/m_{\mathrm{Rb}})$.}
\end{figure}

The experiments which provide the values of the fine structure constant summarized in figure~\ref{Figure.1} can be used in two different ways. On the one hand, they can be seen as a test of QED calculations  of the electron moment anomaly $a_{\mathrm{e}}$. These very difficult calculations have been performed by the group of Kinoshita and Nio. They have recently calculated for the first time the fifth coefficient of  equation~\ref{eqQED} and improved the uncertainty on the fourth one.  For the test of QED only two data are involved: the experimental value of $a_\mathrm{e}(\mathrm{Exp})$ (1159652180.73(28)$\times$10$^{-12}$ [0.24ppb]) achieved by the group of Gabrielse and the one predicted by the theory, $a_{\mathrm{e}}(\mathrm{Theory})$. The latter is computed using $\alpha (h/m_{\mathrm{Rb}})$ as input data. The disagreement between the experimental and theoretical values of $a_{\mathrm{e}}$ is:
\begin{equation}
a_{\mathrm{e}}(\mathrm{Exp})-a_{\mathrm{e}}(\mathrm{Theory})=-1.09(0.83)\times 10^{-12}
\end{equation}
The upper part of  figure \ref{Figure.7} shows the comparison between the current  values of $a_{\mathrm{e}}$. The accuracy of the value of $\alpha (h/m_{\mathrm{Rb}})$  is sufficient to
test the contributions due to the muon and hadrons in the theoretical value of $a_\mathrm{e}$.
The lower part shows the relative contributions to the electron anomaly of the different terms in Equation~\ref{eqQED}.


On the other hand, these experiments provide a way to investigate the impact of the measurements of $\alpha$ and the ratio $h/m$ on the redefinition of the kilogram. Recently the group of H. Muller at the university of California, Berkeley, published a new measurement  of the Compton frequency of the cesium atom  $h/m_{\mathrm{Cs}}c^2$\cite{Muller2013}. Because $c$ has an exact value, this is equivalent to a measurement of $h/m$ that can be compared to ours
(see table~\ref{table3}).
\begin{table*}
\centering
  \begin{andptabular}[0.8\textwidth]{lcc}{The values of the fine structure constant and the ratio $h/m_{\mathrm{u}}$ deduced from the experiments of Harvard, Berkeley and Paris.\label{table3}}%
&$h/m_{\mathrm{u}}$ [m$^2$ s$^{-1}$]&$\alpha^{-1}$\\
Harvard university&3.9903127118(26)$\times$ 10$^{-7}$ [0.65 ppb]& 137.035999173(35) [0.25 ppb]\\
Berkeley university&3.990312738(16)$\times$ 10$^{-7}$ [4.0 ppb]&137.03599872(28) [2.0 ppb] \\
LKB&3.9903127193(50)$\times$ 10$^{-7}$ [1.2 ppb] & 137.035999044(90) [0.66 ppb] \\

  \end{andptabular}
\end{table*}

Reference \cite{Muller2013} highlights the impact of such measurements on the proposed redefinition of the SI of units. The interpretation of the aforementioned work needs to be clarified: in the redefinition planned by the CGPM in 2015, the definition of the second will stay the same and the kilogram will be defined by fixing the value of the Planck constant $h$. This definition will be based on fundamental constants and therefore the resolution of the CGPM explicitly relies on the CODATA for the new definition \cite{CGPM2011}.

The main challenge for the redefinition of the Kilogram, and the main reason why this redefinition has been delayed for several years, is the lack of a reliable link between the microscopic and macroscopic masses. This link is established with a relative uncertainty  of 3 $\times 10^{-8}$ \cite{Williams1998} and  with a large discrepancies between the different methods (watt balances and Avogadro project). One can notice that the recently measured value of the Avogadro constant \cite{Andreas2011}, which is the most accurate input datum for the kilogram redefinition, is midway between the watt-balance values \cite{Becker2012, Mana2012}.

\begin{figure}
\includegraphics[width=\columnwidth]{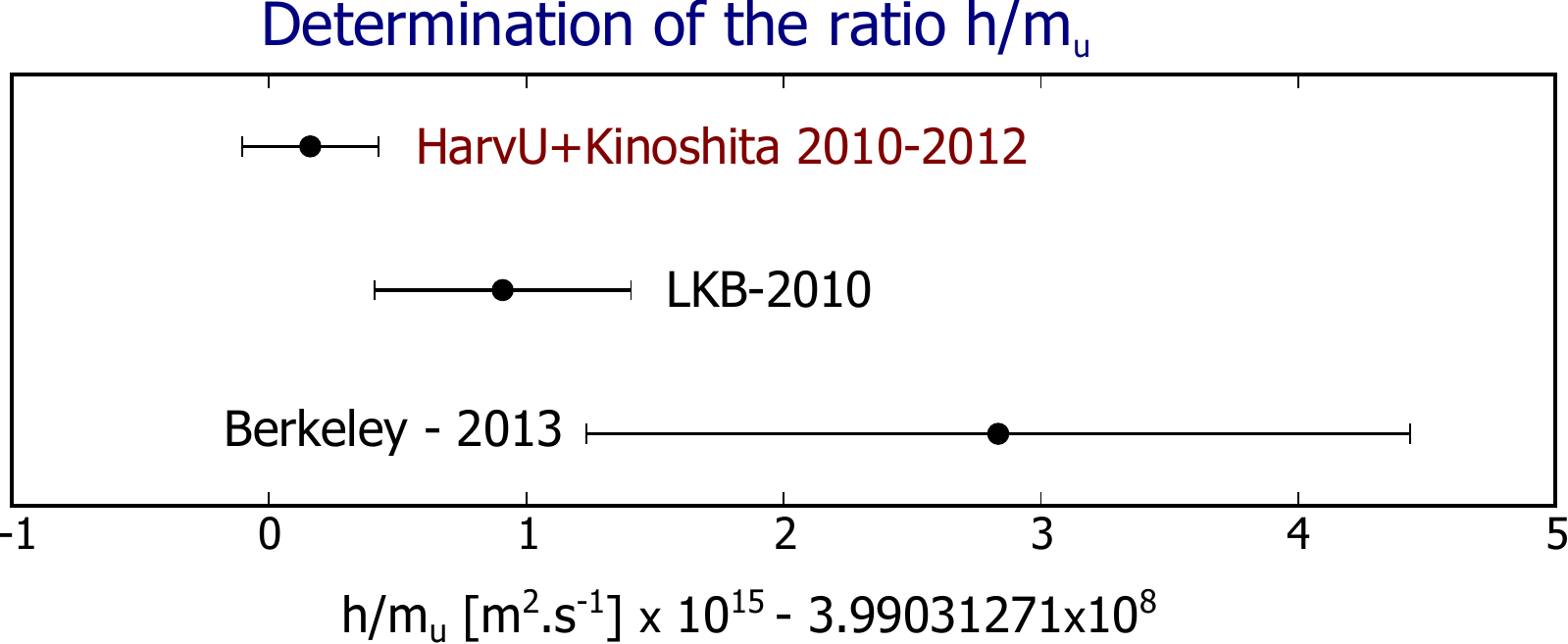}%
  \caption{\label{Figure.8}\col
Determinations of the ratio $h/m_{\mathrm{u}}$ deduced from the measurement of the rubidium recoil and Compton frequency of the cesium atom. The most precise determination comes from the value of the fine structure constant  given by the experimental value of the $a_\mathrm{e}$ measurement and QED calculations.}
\end{figure}

As for the CODATA, the recent measurement of Berkeley is strictly equivalent to an $h/m_{\mathrm{u}}$ measurement. While it contributes to the reliability of the determination of $h/m_{\mathrm{u}}$ by providing a determination below 10$^{-8}$, unfortunately it will not  contribute that much to the CODATA (and therefore to the redefinition of the SI) because its uncertainty is too large (see Figure~\ref{Figure.8}, in which is also included the value obtained from $\alpha(\mathrm{HarvU})$, assuming the exactness of QED calculations). 

The ratio $h/m_{\mathrm{u}}$ will have an important role after the redefinition of the SI. As mentioned in reference \cite{Mohr2008a} \textit{This would yield a value for the mass of the atom in SI units, i.e. kilograms, without making reference
to the prototype kilogram artefact as is now necessary.}
Therefore, there is a strong motivation to continue to improve the uncertainty on $h/m_{\mathrm{u}}$ as much as possible, until competing methods are obviously superior. 

This uncertainty will then be comparable to the ones of the comparison between atomic masses and therefore uncertainties of atomic masses in SI will be the same as in the atomic mass unit (AMU).

The watt balance and $h/m$ measurement are similar in the sense that both measure a ratio between the Planck constant and a mass (or the Compton frequency of a given mass). The watt balance does indeed provide a measurement of the ratio $h/Mg$ ($g$ is the gravity acceleration, $M$ a macroscopic mass). The atomic analogue is closely related to the period $\hbar k /mg$ of Bloch oscillations of atoms in a periodic lattice ($k$ is the wave vector of the Bloch beam). Based on this idea, we have proposed in 2006 to measure the local gravity in the watt balance site using  Bloch oscillations in a quasi-stationary optical lattice. This method gives a possibility  to  realize a link between $h/M$ and $h/m_{\mathrm{u}}$\cite{Biraben2006} .

\section{Conclusion}
In this paper we have presented the details of our recent experimental setup. The fine structure constant is determined with a relative uncertainty  of 6.6$\times$10$^{-10}$. Taking account of the recent improvement of QED calculations, we deduce a theoretical value of the electron moment anomaly. The comparison with the experimental value of $a_{\mathrm{e}}$ realized by the group of Gabrielse at Havard university provides the most stringent test of QED.

In the future we plan to improve the accuracy on $h/m_{\mathrm{Rb}}$ and therefore on $\alpha$, by increasing the sensitivity of the atom interferometer (velocity sensor) and by reducing the systematic effect due to the Gouy phase and the wave-front curvature. A new project is currently in progress in our group. It consists on a new experimental setup based on evaporatively cooled atoms. We plan to implement on this setup an atom interferometer based on large momentum beam splitters \cite{Clade2009LMT}.

This experiment is supported in part by IFRAF (Institut
Francilien de Recherches sur les Atomes Froids), and by
the Agence Nationale pour la Recherche, FISCOM
Project-(ANR-06-BLAN-0192).

\bibliographystyle{andp2012}
\bibliography{publi}

\providecommand{\WileyBibTextsc}{}
\let\textsc\WileyBibTextsc
\providecommand{\othercit}{}
\providecommand{\jr}[1]{#1}
\providecommand{\etal}{~et~al.}


\begin{thebibliography}{[10]}

\bibitem{Bouchendira2011}
 \textsc{R.~Bouchendira},  \textsc{P.~Clad\'e},
  \textsc{S.~Guellati-Kh\'elifa},  \textsc{F.~Nez},  and  \textsc{F.~Biraben}
  \jr{Phys. Rev. Lett.} \textbf{106}(8), 080801 (2011).


\bibitem{Hanneke2008}
 \textsc{D.~Hanneke},  \textsc{S.~Fogwell},  and  \textsc{G.~Gabrielse}
  \jr{Phys. Rev. Lett.} \textbf{100}(12), 120801 (2008).


\bibitem{Aoyama2012}
 \textsc{T.~Aoyama},  \textsc{M.~Hayakawa},  \textsc{T.~Kinoshita},  and
  \textsc{M.~Nio} \jr{Phys. Rev. Lett.} \textbf{109}, 111807 (2012).


\bibitem{Uzan}
 \textsc{J.\,P. Uzan},
 \jr{Living Rev. Relativity} \textbf{14}, 2 (2011).


\bibitem{Mills2011}
 \textsc{I.\,M. Mills},  \textsc{P.\,J. Mohr},  \textsc{T.\,J. Quinn},
  \textsc{B.\,N. Taylor},  and  \textsc{E.\,R. Williams} \jr{Phil. Trans. R.
  Soc. A.} \textbf{369}, 3907 (2011).


\bibitem{Sommerfeld1916}
 \textsc{A.~Sommerfeld},
 \jr{Annalen der Phyisk} \textbf{17}, 1 (1916).


\bibitem{Dirac1916}
 \textsc{P.\,A.\,M. Dirac},
 \jr{Proc. Soc. Lond. A} \textbf{117}, 610 (1928).


\bibitem{Lamb1947}
 \textsc{W.\,L. Jr} and  \textsc{R.\,C. Retherford},
 \jr{Phys. Rev.} \textbf{72}, 241 (1947).


\bibitem{Lamb1950}
 \textsc{W.\,E.\,L. Jr} and  \textsc{R.~Retherford},
 \jr{Phys. Rev.} \textbf{79}, 549 (1950).


\bibitem{Schwinger1948}
 \textsc{J.~Schwinger},
 \jr{Phys. Rev.} \textbf{73}, 416 (1948).


\bibitem{Kusch1948}
 \textsc{P.~Kusch} and  \textsc{H.\,M. Foley},
 \jr{Phys. Rev.} \textbf{74}, 351 (1948).


\bibitem{CODATA98}
 \textsc{P.~Mohr} and  \textsc{B.~Taylor} \jr{Rev. Mod. Phys.} \textbf{72}, 351
  (2000).


\bibitem{Muller2013}
 \textsc{S.\,Y. Lan},  \textsc{P.\,C. Kuan},  \textsc{B.~Estey},
  \textsc{D.~English},  \textsc{J.\,M. Brown},  \textsc{M.\,A. Hohensee},  and
  \textsc{H.~M\"uller} \jr{Science} \textbf{339}, 554--557 (2013).


\bibitem{VanDick}
 \textsc{R.\,V. Dyck},  \textsc{P.~Schwinberg},  and  \textsc{H.~Dehmelt}
  \jr{Phys. Rev. Lett.} \textbf{59}, 26 (1987).


\bibitem{Wicht:02}
 \textsc{A.~Wicht},  \textsc{J.~Hensley},  \textsc{E.~Sarajlic},  and
  \textsc{S.~Chu} \jr{Physica Scripta} \textbf{T102}, 82 (2002).


\bibitem{CODATA2006}
 \textsc{P.\,J. {Mohr}},  \textsc{B.\,N. {Taylor}},  and  \textsc{D.\,B.
  {Newell}} \jr{Reviews of Modern Physics} \textbf{80}, 633--730 (2008).


\bibitem{Udem1997}
 \textsc{T.~Udem},  \textsc{A.~Huber},  \textsc{B.~Gross},
  \textsc{J.~Reichert},  \textsc{M.~Prevedelli},  \textsc{M.~Weitz},  and
  \textsc{T.\,W. H\"ansch} \jr{Phys. Rev. Lett.} \textbf{79}, 2646--2649
  (1997).


\bibitem{Schwob1999}
 \textsc{C.~Schwob},  \textsc{L.~Jozefowski},  \textsc{B.~de~Beauvoir},
  \textsc{L.~Hilico},  \textsc{F.~Nez},  \textsc{L.~Julien},
  \textsc{F.~Biraben},  \textsc{O.~Acef},  \textsc{J.\,J. Zondy},  and
  \textsc{A.~Clairon} \jr{Phys. Rev. Lett.} \textbf{82}, 4960--4963 (1999).


\bibitem{CODATA12}
 \textsc{P.\,J. Mohr},  \textsc{B.\,N. Taylor},  and  \textsc{D.\,B. Newell}
  \jr{Rev. Mod. Phys.} \textbf{84}, 1527--1605 (2012).


\bibitem{Bradley1999}
 \textsc{M.\,P. Bradley},  \textsc{J.\,V. Porto},  \textsc{S.~Rainville},
  \textsc{J.\,K. Thompson},  and  \textsc{D.\,E. Pritchard} \jr{Phys. Rev.
  Lett.} \textbf{83}, 4510--4513 (1999).


\bibitem{Mount2010}
 \textsc{B.\,J. Mount},  \textsc{M.~Redshaw},  and  \textsc{E.\,G. Myers}
  \jr{Phys. Rev. A} \textbf{82}(Oct), 042513 (2010).


\bibitem{Gabrielse2006}
 \textsc{G.~Gabrielse},  \textsc{D.~Hanneke},  \textsc{T.~Kinoshita},
  \textsc{M.~Nio},  and  \textsc{B.~Odom} \jr{Phys. Rev. Lett.}
  \textbf{97}(Jul), 030802 (2006).


\bibitem{CGPM2011}
 \textsc{R.~adopted\,by\,the\,General\,Conference\,on Weights} and
  \textsc{M.~(24th meeting)} (17-21 October 2011).


\bibitem{Borde2005}
 \textsc{C.\,J. Bord\'e},
 \jr{Phil. Trans. R. Soc. A.} \textbf{363}, 2177--2201 (2005).


\bibitem{Mills2006}
 \textsc{I.\,M. {Mills}},  \textsc{P.\,J. {Mohr}},  \textsc{T.\,J. {Quinn}},
  \textsc{B.\,N. {Taylor}},  and  \textsc{E.\,R. {Williams}} \jr{Metrologia}
  \textbf{43}, 227--246 (2006).


\bibitem{Kibble1990}
 \textsc{B.\,P. Kibble},  \textsc{I.\,A. Robinson},  and  \textsc{J.\,H.
  Belliss} \jr{Metrologia} \textbf{27}, 173 (1990).


\bibitem{Steiner2007}
 \textsc{R.\,L. Steiner},  \textsc{R.~Liu},  \textsc{P.T.Olson},  and
  \textsc{D.},
 \jr{IEEE. Trans. Instrum. Meas.} \textbf{56}, 592 (2007).


\bibitem{Williams1998}
 \textsc{E.\,R. Williams},  \textsc{R.\,L. Steiner},  \textsc{D.\,B. Newell},
  and  \textsc{P.\,T. Olsen} \jr{Phys. Rev. Lett.} \textbf{81}, 2404--2407
  (1998).


\bibitem{Andreas2011}
 \textsc{B.~Andreas},  \textsc{Y.~Azuma},  \textsc{G.~Bartl},
  \textsc{P.~Becker},  \textsc{H.~Bettin},  \textsc{M.~Borys},
  \textsc{I.~Busch},  \textsc{M.~Gray},  \textsc{P.~Fuchs},  \textsc{K.~Fujii},
   \textsc{H.~Fujimoto},  \textsc{E.~Kessler},  \textsc{M.~Krumrey},
  \textsc{U.~Kuetgens},  \textsc{N.~Kuramoto},  \textsc{G.~Mana},
  \textsc{P.~Manson},  \textsc{E.~Massa},  \textsc{S.~Mizushima},
  \textsc{A.~Nicolaus},  \textsc{A.~Picard},  \textsc{A.~Pramann},
  \textsc{O.~Rienitz},  \textsc{D.~Schiel},  \textsc{S.~Valkiers},  and
  \textsc{A.~Waseda} \jr{Phys. Rev. Lett.} \textbf{106}(3), 030801 (2011).


\bibitem{Borde1989}
 \textsc{C.\,J. Bord\'e} \jr{Phys. Lett. A} \textbf{140}(1-2), 10 -- 12 (1989).


\bibitem{BenDahan}
 \textsc{M.\,B. Dahan},  \textsc{E.~Peik},  \textsc{J.~Reichel},
  \textsc{Y.~Castin},  and  \textsc{C.~Salomon} \jr{Phys. Rev. Lett.}
  \textbf{76}, 4508 (1996).


\bibitem{Peik}
 \textsc{E.~Peik},  \textsc{M.\,B. Dahan},  \textsc{I.~Bouchoule},
  \textsc{Y.~Castin},  and  \textsc{C.~Salomon} \jr{PRA} \textbf{55}, 2989
  (1997).


\bibitem{Wilkinson1996}
 \textsc{S.\,R. Wilkinson},  \textsc{C.\,F. Bharucha},  \textsc{K.\,W.
  Madison},  \textsc{Q.~Niu},  and  \textsc{M.\,G. Raizen} \jr{Phys. Rev.
  Lett.} \textbf{76}(Jun), 4512--4515 (1996).


\bibitem{Battesti04}
 \textsc{R.~Battesti},  \textsc{P.~Clad\'e},  \textsc{S.~Guellati-Kh\'elifa},
  \textsc{C.~Schwob},  \textsc{B.~G\'emaud},  \textsc{F.~Nez},
  \textsc{L.~Julien},  and  \textsc{F.~Biraben} \jr{Phys. Rev. Lett.}
  \textbf{92}, 253001 (2004).


\bibitem{Cadoret2008}
 \textsc{M.~Cadoret},  \textsc{E.~de~Mirandes},  \textsc{P.~Clad\'{e}},
  \textsc{S.~Guellati-Khelifa},  \textsc{C.~Schwob},  \textsc{F.~Nez},
  \textsc{L.~Julien},  and  \textsc{F.~Biraben} \jr{Phys. Rev. Lett.}
  \textbf{101}(23), 230801 (2008).


\bibitem{Clade2006}
 \textsc{P.~Clad\'{e}},  \textsc{E.~de~Mirandes},  \textsc{M.~Cadoret},
  \textsc{S.~Guellati-Kh\'{e}lifa},  \textsc{C.~Schwob},  \textsc{F.~Nez},
  \textsc{L.~Julien},  and  \textsc{F.~Biraben} \jr{Phys. Rev. A}
  \textbf{74}(5), 052109 (2006).


\bibitem{Baillard2006}
 \textsc{X.~Baillard},  \textsc{A.~Gauguet},  \textsc{S.~Bize},
  \textsc{P.~Lemonde},  \textsc{P.~Laurent},  \textsc{A.~Clairon},  and
  \textsc{P.~Rosenbusch} \jr{Optics Communications} \textbf{266}(2), 609 -- 613
  (2006).


\bibitem{Touahri}
 \textsc{D.~{Touahri}},  \textsc{O.~{Acef}},  \textsc{A.~{Clairon}},
  \textsc{J.\,J. {Zondy}},  \textsc{R.~{Felder}},  \textsc{L.~{Hilico}},
  \textsc{B.~{de Beauvoir}},  \textsc{F.~{Biraben}},  and  \textsc{F.~{Nez}}
  \jr{Optics Communications} \textbf{133}(February), 471--478 (1997).


\bibitem{Witt2007}
 \textsc{T.\,J. Witt} \jr{Metrologia} \textbf{44}(3), 201 (2007).


\bibitem{Becker2012}
 \textsc{P.~Becker} \jr{Contemporary Physics} \textbf{53}(6), 461--479 (2012).


\bibitem{Mana2012}
 \textsc{G.~Mana} and  \textsc{E.~Massa} \jr{Rivista Del Nuovo Cimento}
  \textbf{35}, 353 (2012).


\bibitem{Mohr2008a}
 \textsc{P.\,J. Mohr} \jr{Metrologia} \textbf{45}(2), 129 (2008).


\bibitem{Biraben2006}
 \textsc{F.~Biraben},  \textsc{M.~Cadoret},  \textsc{P.~Clade},
  \textsc{G.~Geneves},  \textsc{P.~Gournay},  \textsc{S.~Guellati-Khelifa},
  \textsc{L.~Julien},  \textsc{P.~Juncar},  \textsc{E.~de~Mirandes},  and
  \textsc{F.~Nez} \jr{METROLOGIA} \textbf{43}(6), L47--L50 (2006).


\bibitem{Clade2009LMT}
 \textsc{P.~Clad\'e},  \textsc{S.~Guellati-Kh\'{e}lifa},  \textsc{F.~Nez},  and
   \textsc{F.~Biraben} \jr{Phys. Rev. L} \textbf{102}(24), 240402 (2009).


\end{thebibliography}

\end{document}